\documentclass[letterpaper,twocolumn,10pt]{article}
\usepackage{usenix2019}

\usepackage{tikz}
\usepackage{amsmath}

\usepackage{filecontents}

\begin{document}

\date{}

\title{\Large \bf Security for People with Mental Illness in Telehealth Systems: A Proposal}

\author{
{\rm Helen Jiang}\\
Independent (affiliated with Georgia Institute of Technology)
} 

\maketitle

\begin{abstract}
A mental health crisis is looming large, and needs to be addressed. But across age groups, even just in the United States, more than 50\% of people with any mental illness (AMI) did not seek or receive any service or treatment\cite{stats:nih}. The proliferation of telehealth and telepsychiatry tools and systems\cite{telehealth:apa,telehealth:psychiatry} can help address this crisis, but outside of traditional regulatory aspects on privacy, e.g. Health Insurance Portability and Accountability Act (HIPPA), there does not seem to be enough attention on the security needs, concerns, or user experience of people with AMI using those telehealth systems. 

In this text, I try to explore some priority security properties for telehealth systems used by people with AMI for mental heath services (MHS). I will also suggest some key steps in a proposed process for designing and building security mechanisms into such systems, so that security is accessible and usable to patients with AMI, and these systems can achieve their goals of ameliorate this mental health crisis.f 

\end{abstract}

\section{Introduction}
Mental health issues are prevalent around all of us, and the scale is staggering. Within the United States alone, in 2017, 46.6 million adults had a mental illness, 49.5\% of adolescents had any mental disorder, and 10.6 million adults seriously considered suicide\cite{stats:nih, stats:shamsa}. Estimates are that 50\% of mental illness begins by age 14, and 75\% by 24, while suicide is the third biggest cause of death for age group 10 -- 24, among whom 90\% had underlying mental illness\cite{mentalillness:apa,mentalillness:pnas}. Telehealth and telepsychiatry tools and systems have been developed with the hope to help address this crisis, and while they all must comply to HIPPA, there is a missing dimension: psychologically acceptable security to people with AMI. 

The ``psychological acceptability'' principle is identified as ``usable'' in 1975\cite{saltzer1975protection}, but it wasn't until the 1990s, that ``usable security'' started to get its due attention\cite{og:zurkoandsimon1996, usable_sec_gatech_payne2008brief}. Moreover, the audience of ``psychological acceptability'' is open and wide: \emph{to whom}, or \emph{to what audience} are the security measures and mechanisms psychologically acceptable? What if the psychological or mental state of the audience is impaired, or the audience has mental disorders? 

This question is open, wide, and more importantly, tricky. While it is relatively easy to diagnose and notice cognitive impairment and neurocognitive disorders as they manifest in domains such as attention, recognition, and language~\cite{diagnose:cognitive:cdc,diagnose:cognitive:nih,diagnose:cognitive:dsm5}, the vast majority of people with mental illness keep functioning in daily lives\cite{mentalillness:apa:functioning}. What is even trickier, is that mental disorder may eventually turn to affect cognition and behaviors, as Diagnostic and Statistical Manual of Mental Disorders (DSM, latest edition DSM-5) defines a mental disorder as ``...\emph{a syndrome characterized by clinically significant disturbance in an individual’s cognition, emotion regulation, or behaviors}...''\cite{dsm5:def}. How might we build security into telehealth systems, which would be relied upon by many with mental illness, who a diverse and complex, but under-served and usually invisible user base? This is a question worth asking and solving. In this work, I will propose some priority properties of security in telehealth systems used for MHS, and suggest some key steps in a process when building usable and secure telehealth systems for people with AMI. Here I will adapt \cite{def:telehealth}'s definition of telehealth, to better suit the MHS context. While it is still fundamentally ''\emph{the use of electronic information and telecommunication technologies to support long-distance clinical health care} etc.,'' providers of MHS via telehealth need not to be only human --- they can be automated, interactive agents such as social bots, e.g. conversational agents (colloquially ``chatbots'').

\section{Related Work}
For security and usable security, much has been written and researched. However, even though ``psychological acceptability'' to users was proposed as a key principle for security, it only started getting attention much later. Meanwhile, security measures keep confusing users\cite{og:zurkoandsimon1996, notenemy:adams1999users, cantencrypt:whitten1999johnny, stillcantencrypt:sheng2006johnny, recipro:chiasson2007even}. Moreover, the ``psychological acceptability'' principle is often doubted as incompatible with the goal of ``security''\cite{og:zurkoandsimon1996, dewitt2006usable, smetters2007usable, theofanos2020usable, soup_sec_comp, usesec:balfanz2004search, usablesec:yee2004aligning, hci:patrick2003hci}, and usable security is still a small community compared to other areas of security research. Also, as \cite{incluog:3rdwave:wang2017third} points out, usable security is designed with the general population in mind, and may leave out specific vulnerable groups that are under-served. This leaves us not a deep foundation to work with, when we consider building psychologically acceptable security, for those whose mental state may suffer from disorders or illness, and into systems that many of them may rely on to get treatment and services. 

But more recently, usable security for vulnerable groups is getting more attention,  especially on older adults and the visually impaired, thanks to works such as\cite{older:munteanu2015improving,older:nicholson2019if,older:nicholson2019if,older:watch:caine2011digiswitch,vis:ahmed2016addressing,vis:dosono2015m,vis:haque2013secure,vis:rector2015exploring,vis:vatavu2017improving,vis:ye2014current,visprivacy:ahmed2015privacy} from both usable security and human-computer interaction (HCI) communities. On cognitively impaired users, \cite{older:study:mentis2019upside} did an excellent user study on older adults with mild congnitive impairments and their online security behaviors on sharing passwords and identity information, and while it discussed risks options and gave examples on access and control, it stopped short of providing security-specific suggestions, properties, principles, or processes. \cite{neuro:ma2013investigating} investigated behaviors of certain cognitively impaired users on authentication methods, but only in a simulated e-commerce setting, not in telehealth systems used for MHS: such systems hold a lot more sensitive information and interactions from and about the users. 

On the other hand, the rise of telehealth systems for mental health (e.g. social bots and online therapy) has prompted active research from the HCI community and health researchers \cite{chi20_digi_trad_depres:system,chi20:bendelin2011experiences,chi20:andersson2009internet,chi20:andersson2018long,chi20:beattie2009primary, chi20:bot:bickmore2010maintaining,chi20:bot:bickmore2010response,chi20:bot:ferrara2016rise,chi20:bot:fitzpatrick2017delivering,chi20:bot:grimme2017social,chi20:bot:ly2017fully,chi20:bot:techprivacy:yarosh2013shifting,chi20:botboss,chi20:humaninfra,arttherapy:cornejo2016vulnerability}, but their primary focus is on users' experience in treatment, their system interaction experiences, and effects of therapy, and did not give much consideration to security --- the psychologically acceptable type of security --- to users with AMI. \cite{chi20:bot:techprivacy:yarosh2013shifting} uncovered the general ``tensions with technology'' among members of several peer-support therapy groups, and while it identified ``anonymity, identity, access'' as parts of the tensions, it focuses on user experiences and participation, and did not address inclusive security. 

For telehealth systems, at least in the United States, the federal and state guideline on their security mechanisms are inconsistent\cite{telehealth:nejm:tuckson2017telehealth}, and while medical professionals have proposed\cite{telehealth:daniel2015policy,telehealth:fogel2017survey,telehealth:lerouge2013crossing,telehealth:luxton2012mhealth,tele:privacy:older:vines2013making} measures to improve security in telehealth systems, those measures are largely targeting security concerns of medical institutions\cite{telehealth:nih:langer2017cyber} and tend to be more policy- and administration-oriented than patient-focused.

\section{A Proposal on Properties \& Process}
These properties are by no means exhaustive or authoritative. They are my early explorations into making security methods, mechanisms, and designs that are easily accessible and usable by people with AMI. Most, if not all of these properties have been discussed in general computer science and security literature before. However, putting them into the context for providing security to people with AMI accessing MHS via telehealth systems, places them in high priority positions for the practical design and implementation of those telehealth systems used for MHS. 

	\subsection{Some Priority Properties} \label{sec:priorityprops}
			\begin{description}
				\item[Trust-inducing] As \cite{chi20:bot:techprivacy:yarosh2013shifting} illustrates, distrust of technology (e.g. videochats, social networks, cloud data storage) forms the basis of tension with technology in several peer-support therapy groups, and older adults with cognitive impairment are concerned about their privacy when doing art therapy online too\cite{arttherapy:cornejo2016vulnerability}. With this distrust of general technology already in the minds of users who seek MHS, it is extremely important for telehealth systems --- which handles more sensitive data, content, and interactions than general technology platforms and services --- to earn and induce trust from users to ensure and encourage adoption and usage. But security measures --- good security measures --- could easily confuse or mislead users\cite{og:zurkoandsimon1996, notenemy:adams1999users, cantencrypt:whitten1999johnny, stillcantencrypt:sheng2006johnny, recipro:chiasson2007even,johnny:ruoti2013confused}, and when confusion and misdirection in telehealth systems happen to users with AMI seeking MHS, these users may withdraw altogether from using telehealth systems, and their withdrawal may have larger and more detrimental effects on their well-being than the general population withdrawing from telehealth systems.  
				
				\item[Robust] By definition\cite{dsm5:def}, mental disorders can disrupt a person’s behaviors and cognition, and such disturbances may not be within considerations of user models or behavioral expectations of general usable security, which has the general population in mind. Therefore, it would be wise to expect more errors, faults, and unexpected inputs or behaviors from users with AMI, and let this expectation lead to the building of more robustness into security measures of telehealth systems used for MHS. Robustness is especially important to telehealth systems whose providers of MHS are automated interactive agents, which are more affordable and accessible than human MHS providers. They are very likely to be relied upon more by people with AMI whose socio-economic status and life circumstances could not afford them to human MHS providers, and without more robustness, these automated interactive agents may deny access to care to users needing the care most. 
				
				\item[Cooperative] MHS is naturally a cooperative process, with interactions between patients, providers, other caregivers, and peers active and evolving in the MHS process. People with AMI may have delegated certain powers to caregivers other than their providers, and may also seek MHS with others in scenarios such as family therapy and peer-support group therapies. In the light of this situation, it sounds sensible to consider making certain security mechanisms and features cooperative. But this cooperative nature of MHS may pose legal challenges for security: take the U.S. for example, a shared password, even a voluntarily shared one, counts as unauthorized access by the Computer Fraud and Abuse Act, despite technology policy organizations' persistent activism\cite{arttherapy:cornejo2016vulnerability}. In spite of this challenge, there are times when such sharing and other cooperative procedures are necessary during one's life span. Telehealth systems used for MHS should proactively consider building security measures that address the roles and functions of caregivers, peers, partners, etc., and in the process of delivering MHS, to actively communicate the system's cooperative aspect of security to users.
				
				\item[Functional] This may sound silly and obvious, but given the infamous security-functionality trade-off\cite{og:zurkoandsimon1996, usesec:kainda2010security, to:albrechtsen2010improving, to:hagen2009effects, to:navarro2005approximating} even for the general population, it is worth emphasizing functionality when designing security to include people with AMI, who need MHS provided via telehealth systems more than the general population. Usable security measures should not become obstacles for people with AMI when they access telehealth systems' care-providing capabilities, and ideally, should not degrade their general user experiences either. 
				
				\item[Fail gracefully] Security methods, however carefully designed and built, may still fail to guide users towards the right path of actions, but a graceful failure can help lead users to the right path next time. When the telehealth systems' security mechanism fails to elicit the right actions from users with AMI, how should the system respond, so next time users can do the right things? Given mental disorders could affect cognition and behavior, would users with AMI react differently to security warnings and failure messages as the general population do? If so, how different and in what aspects? What post-failure contents or actions, educational or otherwise, from the telehealth system can help users do the right thing the next time? These are important questions to consider, when designing and building security mechanisms in telehealth systems to provide MHS to users with AMI.
			
	       \end{description}
	
	\subsection{Process: Some Key Steps}
Based on preceding discussion, it is now helpful to start contemplating the practical aspect of building those priority security properties into telehealth systems used for MHS. I propose some key steps to make security inclusive to users with AMI, and invite the wider usable security community to further contribute to, discuss, debate, utilize, improve, and architect processes that would engineer more inclusive security mechanisms in telehealth systems for people with AMI. 

		\begin{description}			
			\item[Build inclusive mental models]\cite{mentalmodel:inclusec:houser2017formal} advocates for formal methods when building mental models for inclusive security's user models, and in practice, it might be realistic to first build mental models for the security concerns and potential ranges of behaviors for a sub-population of people with AMI, whom the telehealth system targets. These users may differ and diverge from the general population on how they perceive, expect, and manage security and risk in telehealth systems, and their behaviors can range wider than the general population when it comes to use security methods. Including these considerations about the concerns and behaviors of people with AMI, would be the first step towards inclusive security in telehealth systems providing MHS. 
			
			An example: young adults with attention-deficit/hyperactivity disorder (ADHD) who seek conversational therapy from an automated agent (``chatbot'') online. While CAPTCHA methods could help protect the chatbot against spams, users with ADHD --- who generally have shorter attention spans than a general population\cite{adhd:nimh} --- may be more likely to abandon efforts to go through CAPTCHAs, especially when there are multiple and come one after another in various forms (e.g. text, sound, or images recognition). They may even leave with the belief that such a security mechanism is set up to trick them or deny them access. Here, an inclusive mental model of user behaviors does not only help users access MHS, but also encourage creative solutions. For instance, instead of human-recognition-based CAPTCHAs which may require sustained non-interactive attention, might a 30-second ``trial conversation'' with the chatbot to decide ``human-or-not '' be useful?
			
			\item[Incorporate clinical providers' inputs] Continuing on the previous point, understanding and modeling behaviors and concerns of people with AMI, cannot be done in a vacuum or in an armchair checking off criteria in DSM-5\footnote{DSM provides a common framework for describing psychopathology, and it is strongest in its reliability. Meanwhile, there has always been controversies around DSM about its criteria, categorization, characterization, and clinical validity\cite{diagnose:fp:wakefield2016diagnostic,dsm5:controversy:blog}.}. To build realistic user models and their mental models of security, consulting clinical practitioners on what ranges of behaviors to expect from patients with AMI seeking MHS, would be helpful. Practically speaking, incorporating clinicians views is the next best available approach, short of actual target user research and interviews: in environments where AMI is stigmatized, people with AMI may not be willing to disclose their conditions to researchers, let alone wanting to be interviewed, observed, and studied for their behaviors using telehealth systems. Whenever pragmatic and viable, user research is still the preferred and best method of research and user modeling, but in its absence, a good alternative would be to seek clinicians' inputs on the online behavioral patterns they have observed in their patients while providing MHS, and then to extract possible mental models of security of people with AMI. 
						
			\item[Consider cooperative situations in security]As described in section \ref{sec:priorityprops}, MHS is cooperative by nature, and in the context of telehealth systems in MHS, ``cooperations'' can happen between human beings and automated interactive agents. Hence, building technical security mechanisms and writing user-facing security communications (e.g. user agreements) that would allow security to be cooperative in certain MHS contexts, would be one important way that telehealth systems can make security more inclusive for people with AMI seeking MHS. 
			
			\item[Define boundaries of cooperation]While cooperation is important in MHS, it is also crucial to \emph{not} idealize it in security settings, especially for security in MHS. Boundary violations between providers and patients seeking MHS, and exploitation of those patients from their caregivers and peers\cite{boundary:gutheil1993concept,boundry:smith1995patient,boundary:epstein1990exploitation,boundary:gabbard1996lessons} can and do happen in off-line settings. When MHS is provided online via telehealth systems, the boundary and exploitation problems only get more complex with all the security methods and mechanisms in place, so that people with AMI can access MHS. When designing and building cooperative aspects of security in telehealth systems used for MHS, we should also consider drawing boundaries between those principal users with AMI who seek MHS, and the extent to which their providers, caregivers, and peers could influence and change the security decisions, settings, and behaviors of these principal users. 
			
			\item[Tailor communications] This is a logical conclusion following all preceding points. What works well for general populations in general systems to enhance security --- e.g. pop-up warnings, color-coded buttons, push notifications, conventional user interfaces etc. --- may or may not work well with people with AMI seeking MHS via telehealth system. What is more, certain users with AMI may also be cognitively impaired, posing even more challenges. Having built inclusive and diverse user models, it makes sense to implement an inclusive and diverse set of communications paradigms and tools --- ranging from warning message re-writes to user interfaces customization --- so that each user model is accounted for when encountering security mechanisms in telehealth systems.
												
			\item[Evaluate failures]During early stage development and small-scale user trials, it is important to document and evaluate failures, where security mechanisms and methods fail to lead people with AMI to the right action paths or accomplish their purposes. Those failures may be due to incorrect assumptions about user behaviors, insufficient robustness, unclear warnings and failures, incorrect implementations, or a variety of other reasons. Regardless of the reason, learning from those failures and how they might have become hurdles to actual patients using telehealth systems for MHS, or in fact have motivated insecure behaviors, would be very valuable lessons to inform future inclusive security designs and implementations for people with AMI not only in telehealth systems, but also in general technology services and products. 
			
		\end{description}

\section{Future Work}
This is still early stage work, and the landscape of building inclusive security for people with AMI is an open field with many open problems and solutions. These properties and steps are a suggestion, an invitation, and an encouragement for the usable security and inclusive security community to examine, understand, and build towards the security needs of people with AMI. Telehealth systems used for MHS are the most obvious first target, and the lessons we will have learned here could inspire, expand into, transferred to, and be adapted in security designs and mechanisms of other technology services and products to include more under-served groups. 

One direction that I might consider to carry this research forward, is to observe, research, conduct interviews on, and evaluate security behaviors in a specific sub-population of people with AMI, for example, adults with ADHD, who use telehealth systems to access MHS provided by automated, interactive agents. Another possible direction is to evaluate current security solutions in a popular telehealth systems used for MHS, and how these solutions are inclusive or exclusive towards certain people with AMI. 

\section{Conclusion}
With mental health issues prevalent in our societies and telehealth systems proliferating, more people with mental illness are and may start seeking mental health services via telehealth systems. However, there does not seem to be enough discussion on how, and if, security considerations in telehealth systems are including people with mental illness, who may very likely have large cognitive and behavioral deviations from the general population, for whom many security mechanisms are designed and built. 

In this text, I shared some security properties that should be prioritized when building telehealth systems for people with mental illness to access mental heath services. I also suggested some key steps when designing and building security mechanisms and experiences into those systems. I hope that readers can take away not only awareness in the security needs of people with mental illness, but also insights on how the usable security community can start contributing to this important but under-served population that deserve our attention when building usable and inclusive security. 

\bibliographystyle{plain}
\bibliography{hj_wisp_soups2020}

\end{document}